\documentclass[useAMS,usenatbib]{mn2e}
\usepackage{amsmath,graphicx,psfig}

\def\arcmin{$^{\prime}$}
\def\arcsec{$^{\prime\prime}$}
\def\micron{$\rm{\umu}$m}
\def\microns{$\rm{\umu}$m}
\def\xflux{ergs s$^{-1}$ cm$^{-2}$}
\def\xlum{ergs s$^{-1}$}

\title[The contribution of AGN to the submillimetre population]{The contribution of AGN to the submillimetre population} \author[M. D. Hill \& T. Shanks] {Michael D. Hill$^{1}$\thanks{e-mail:
m.d.hill@durham.ac.uk} \& Tom Shanks$^{1}$ \\ $^{1}$Dept.\ of Physics, Science Laboratories, Durham University, Durham, DH1 3LE, UK}

\begin{document}

\pagerange{\pageref{firstpage}--\pageref{lastpage}} \pubyear{2010}

\maketitle

\label{firstpage}

\begin{abstract} 

We find that X-ray sources in the Extended \emph{Chandra} Deep Field South are strongly spatially correlated with LABOCA 870 micron sources. We investigate the dependence of this correlation on X-ray flux, hardness ratio and column density, finding that specifically faint and absorbed X-ray sources are significant sub-mm emitters. In the X-ray source redshift subsample we confirm the previous result that higher luminosity sources ($L_X>10^{44}$ ergs s$^{-1}$) have greater 870\micron\ fluxes but we also find that this subsample selects against absorbed sources, faint in X-ray flux. Overall, we find that X-ray sources contribute $1.5\pm0.1$ Jy deg$^{-2}$ to the sub-mm background, $\approx3$\% of the total, in agreement with the prediction of an obscured AGN model which also gives a reasonable fit to the bright sub-mm source counts. This non-unified model also suggests that when Compton-thick, X-ray-undetected sources are included, then the fractional AGN contribution to the sub-mm background would rise from $\approx3$\% to a total of 25--40\%, although in a unified model the AGN contribution would only reach $\approx13$\%, because the sub-mm flux of the X-ray sources is then more representative of the whole AGN population. Measurements of the dependence of sub-mm flux on X-ray flux, luminosity and column density all agree well with the predictions of the non-unified AGN model. Heavily absorbed, X-ray-undetected AGN could explain the further cross-correlation we find between sub-mm sources and $z>0.5$ red galaxies. We conclude that sub-mm galaxies may contain the long-sought absorbed AGN population needed to explain the X-ray background.

\end{abstract}

\begin{keywords} submillimetre -- galaxies: high-redshift -- quasars: general -- X-rays: galaxies\end{keywords}

\section{Introduction}

It is now more than a decade since the first blank field surveys with the SCUBA instrument (Holland et al., 1999) on the James Clerk Maxwell Telescope revealed the existence of large numbers of highly luminous sources for which a significant fraction of the total bolometric output is in the far-infrared and submillimetre (e.g.\ Smail et al., 1997; Barger et al, 1998; Hughes et al., 1998). These objects were quickly found to be at high redshifts and to be heavily obscured by dust (for a review see Blain et al., 2002). There is now evidence that they host considerable star formation as well as concurrent AGN activity (Alexander et al., 2005). Although many authors have concluded that the former process dominates the bolometric luminosity of sub-mm galaxies, the contribution of AGN nevertheless remains unclear.

Obscured quasars are the primary candidates to explain the ``missing'' hard X-ray background (e.g. Worsley et al., 2005). Since these highly absorbed sources are likely to be dust-rich objects, they would be expected to have substantial luminosities in the infrared where the reprocessed light is emitted. The wavelength regime in which this emission peaks will depend upon the typical temperature of the obscuring dust. If it is cool enough, sub-mm galaxies are viable candidates for this long-sought obscured quasar population. Obscured AGN models have been shown to give a reasonable fit to the bright end of the sub-mm source counts (Gunn \& Shanks, 1999), while star-forming galaxies are expected make the dominant contribution at fainter fluxes (Busswell \& Shanks, 2001). 

Many authors have presented sub-mm and millimetre observations of high-$z$ QSOs (e.g. Andreani et al., 1993; Chini \& Kr\"{u}gel, 1994; McMahon et al., 1994; Omont et al., 1996, 2001, 2003; Hughes et al., 1997; Priddey et al., 2003; Petric et al., 2006; Coppin et al., 2008a), however the issue of whether the active nucleus is ultimately responsible for the high luminosities at long wavelengths remains a largely open question. Observational analyses are often unable to exclude either a starburst-dominated or an AGN-dominated system (e.g. Priddey \& McMahon, 2001; Petric et al., 2006).

One of the key tenets of the unified AGN model (Antonucci, 1993) is that the nucleus is surrounded by dust with a toroidal geometry, such that different viewing angles provide access to different regions, thus giving rise to the separate classifications of narrow- and broad-line AGN. This dust is irradiated by the active nucleus and since the luminosities of these sources are very high the surrounding dust can be heated to high temperatures. If the sub-mm emission results from dust irradiated by the AGN instead of star-formation, it must lie far enough from the central engine to maintain a cool ($\sim30$K) temperature. Simple torus models have shown that this is possible -- the extent of the dusty torus can reach the kiloparsec scale, and dust at these large radii can be cool enough to produce sub-mm/millimetre emission consistent with observed spectra (Granato \& Danese, 1994; Andreani et al., 1999; Kuraszkiewicz et al., 2003).

In recent years, sub-mm observations of AGN have somewhat called into question the unified AGN model, with studies such as that by Page et al.\ (2004) indicating that more X-ray absorbed AGN may be much brighter in the sub-mm than unabsorbed AGN. Such a dichotomy is not easily explained by the unified model, since sub-mm emission is not absorbed by dust or gas and so should not be affected by viewing angle. Instead, absorbed and unabsorbed AGN may form an evolutionary sequence (Fabian, 1999).

In this paper we use the unprecedented combination of depth and width in the sub-mm and X-ray data of the Extended \emph{Chandra} Deep Field South to present new tests of predictions of models for the AGN contribution to the sub-mm number counts and background. We shall mainly be using cross-correlation and stacking techniques which avoid the need to identify individual sub-mm sources with individual X-ray or optical sources.  Such identifications are always difficult when the sub-mm resolution is poor and here we restrict ourselves to statistical measurements of the properties of sub-mm sources near X-ray and optical sources.

The first most basic test of such models is that AGN X-ray sources should cross-correlate on the sky with sub-mm sources. Previous results have been ambiguous -- Almaini et al.\ (2003) measured a positive correlation out to large angular separation ($\theta\approx2$\arcmin), while Borys et al.\ (2004) found that although a large fraction of sub-mm sources had coincident X-ray sources ($\theta<7$\arcsec), there was no correlation out to larger separations. 

The X-ray and sub-mm data can also test unified versus non-unified pictures for the AGN absorption (Page et al., 2004). The most heavily absorbed AGN may only be detected via the host galaxy at optical/IR wavelengths; we shall therefore also look for cross-correlations between sub-mm sources and normal, high redshift galaxies and compare any excess to AGN model predictions.

In \S\ref{s-data} we briefly review the main datasets used in our investigation and in \S\ref{s-method} we describe the cross-correlation techniques we have employed. In \S\ref{s-smx} we present the results of our correlations of sub-mm sources with X-ray sources and in \S\ref{s-optcorr} we show the results of our correlation analyses for optical sources. In \S\ref{s-ebl} we consider the possible contribution of AGN and galaxies to the sub-mm number counts and extragalactic background. Finally, in \S\ref{s-disc} and \S\ref{s-summ} we discuss and summarise our results.

\begin{figure}
\includegraphics[width=80.mm]{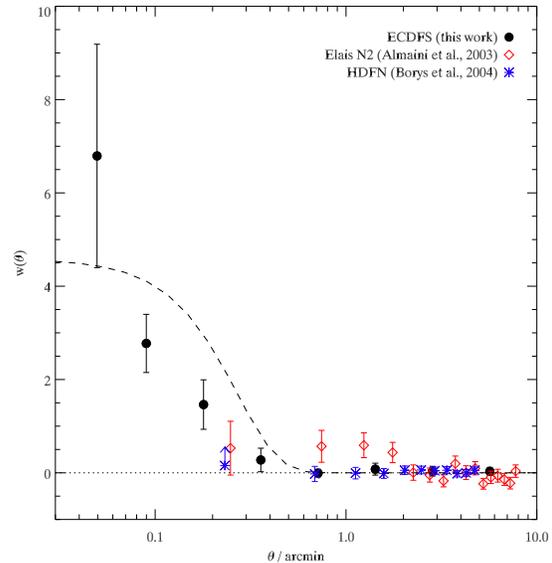}
\caption{Cross-correlation of sub-mm sources with X-ray sources showing a strong correlation between the two populations. An approximation of the LABOCA beam profile is indicated by the black dashed line; the correlation peak lies within the beam profile, showing no positive correlation out to larger separations. This is indicative of the detection of source counterparts rather than associated objects tracing the same structure. For comparison we show previous results by other authors: Almaini et al.\ (2003) detect an extended correlation at large separations, in contrast to our results and those of Borys et al.\ (2004). Borys et al.\ excluded pairs at low separation from their analysis, so we show the first point as a lower limit.} 
\label{f-smallx}
\end{figure}

\begin{figure*}
\includegraphics[width=175.mm]{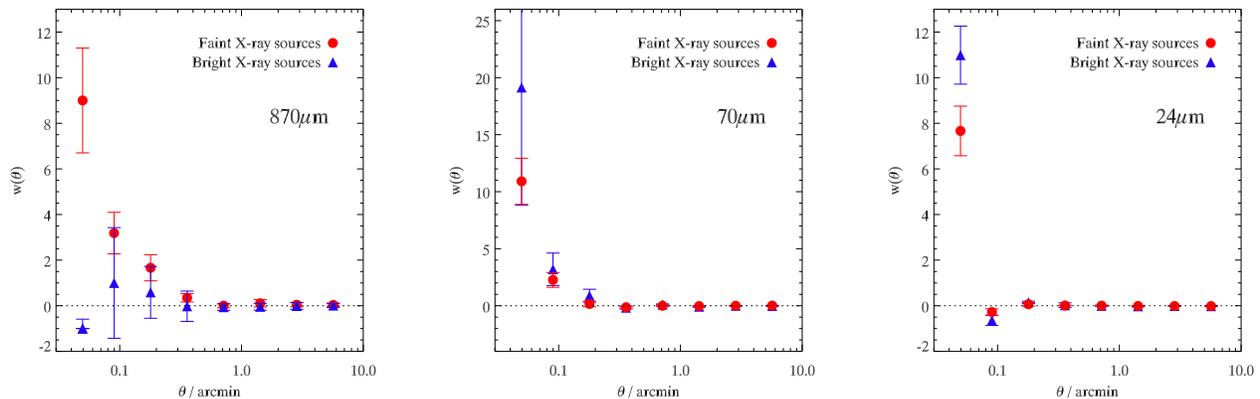}
\caption{Cross-correlation of (\emph{a}) sub-mm, (\emph{b}) 70\micron\ and (\emph{c}) 24\micron\ sources with X-ray sources, with jackknife errors. Red circles represent the the correlations for faint X-ray sources ($S_{0.5-2.0}<1.5 \times 10^{-15}$\xflux), blue triangles show that for bright sources. At 24\micron\ and 70\micron, bright X-ray sources show a stronger correlation, but at 870\micron\ this trend appears to be reversed. Note that the vertical scale is different for the central panel.} 
\label{f-ftbrt}
\end{figure*}

\begin{table*}
\caption{Detected and expected numbers of X-ray--submillimetre pairs at separations of $\theta < 30$\arcsec for all X-ray sources in the ECDFS and separately for those with bright and faint X-ray fluxes. The ``expected" number of pairs is that which would arise from a random distribution of sources, while the detected number is that found from the real X-ray sources. In the first two bins we find 58 faint X-ray--submillimetre pairs, compared to an expectation of 23.8 pairs --- a $7.0\sigma$ excess assuming Poisson statistics, which we have shown to be reliable at these angular separations. The bright sources show a far less significant excess, with 9 pairs in the first two bins compared to an expected 5.5.}
\label{t-smxray}
\begin{tabular}{|c|c|c|c|c|c|c|}
\hline
$\theta$ / arcsec & \multicolumn{2}{c}{All X-ray sources} & \multicolumn{2}{c}{Faint X-ray sources} & \multicolumn{2}{c}{Bright X-ray sources} \\
 & Detected & Expected & Detected & Expected & Detected & Expected \\
\hline
$0-10$ & 30 & 7.5 &27 & 6.0 & 3 & 1.3\\
$10-20$ & 37 & 22.0 & 31 & 17.8 & 6 & 4.2\\
$20-30$ & 40 & 37.4 & 34 & 30.0 & 6 & 7.4\\
\hline
\end{tabular}
\end{table*}

\section{Data} \label{s-data}

In this work we have made use of multiwavelength data from the Extended \emph{Chandra} Deep Field South (ECDFS), which covers $\approx30 \times 30$ arcmin$^2$. The ECDFS comprises four contiguous fields, each with $\approx250$ks \emph{Chandra} exposures, arranged in a $2\times2$ grid centred on the 1Ms \emph{Chandra} Deep Field South (CDFS). We use the X-ray point source catalogue presented by Lehmer et al.\ (2005), who detect 762 X-ray point sources in the ECDFS. They give accurate positional measurements with a median uncertainty of $0.\!\!^{\prime\prime}35$ and provide soft band (0.5--2keV), hard band (2--8keV) and full band (0.5--8keV) fluxes. 

\emph{Chandra}'s increased off-axis PSF means that the sensitivity in each of the four ECDFS frames is worst around the edges  -- this includes the centre of the ECDFS where the four individual frames meet. This central area was covered by the original CDFS observations; we therefore combine the catalogue of Lehmer et al.\ (2005) with the CDFS catalogue of Giacconi et al.\ (2002). We match sources to avoid duplication and include only those CDFS sources with soft-band fluxes of $S_{0.5-2.0}\ge 1\times10^{-16}$\xflux, such that the depth of the central region is the same as that of the extended field. This yields a final ECDFS catalogue of 852 X-ray sources.

We make use of a sample of X-ray sources with confirmed spectroscopic redshifts. These are taken from a survey of the central CDFS area by Tozzi et al.\ (2006) and a wider survey of the full ECDFS by Treister et al.\ (2009). Objects from Treister et al.'s catalogue lying within 3\arcsec\ of an object in Tozzi et al.'s catalogue are excluded. This yielded a spectroscopic redshift sample of 277 X-ray sources. We also supplement this with 114 photometric redshifts from Tozzi et al.'s (2006) catalogue.

The ECDFS was observed at 870\microns\ with the Large APEX Bolometer Camera (LABOCA; Siringo et al., 2009) on the APEX telescope (G\"usten et al., 2006) as part of the LABOCA ECDFS sub-mm Survey (LESS; Wei\ss\ et al., 2009). With $\approx 200$ hrs total on-source integration time these observations reached a uniform noise level of 1.2 mJy beam$^{-1}$ over the full $30^{\prime}\times30^{\prime}$ area. We make use of the resulting beam-smoothed sub-mm map as well as the catalogue of 126 sources detected at $\ge3.7\sigma$ (Wei\ss\ et al., 2009).

At 24\microns\ and 70\microns\ we use public imaging from the Far-Infrared Deep Extragalactic Legacy Survey (FIDEL; Dickinson et al., 2007) using the Multiband Imaging Photometer for Spitzer (MIPS; Rieke et al., 2004). The SExtractor source extraction software was used to obtain source lists from these images.

The ECDFS is one of four fields observed in the optical/near-IR as part of the Multiwavelength Survey by Yale-Chile (MUSYC). We make use of their publicly released BVR-selected catalogue (Gawiser et al., 2006) which has $\sim$85,000 objects and reaches a depth of $R<26.4$.

\section{Methodology} \label{s-method}

We perform statistical analyses of ECDFS sources in the X-ray, sub-mm, optical and infrared, making use of the angular two-point correlation function, $w(\theta)$, defined as:

\begin{equation}
w(\theta)=\frac{DD}{DR} -1
\end{equation}

\noindent where $DD$ and $DR$ are the number of data--data and data--random pairs at a given angular separation $\theta$. 

We also use another formulation of the correlation function which is equivalent to a stacking analysis. Here, we cross-correlate the objects in a source list with each individual pixel of a flux map in order to measure the excess of flux associated with the objects in the list. We find the average pixel value within annuli of increasing radius, having subtracted the mean flux. In this case $w(\theta)$ is defined as:

\begin{equation}
w(\theta)=\frac{\displaystyle\sum_{i=1}^{n_{\theta}}(f_i(\theta)-\bar{f} ) }{n_{\theta}}
\end{equation}

\noindent where $n_{\theta}$ is the number of pixels in a given $\theta$-bin, $f_i$ is the value of the $i^{\rm{th}}$ pixel in that bin and $\bar{f}$ is the average pixel value.

We measure the uncertainties on our cross-correlation results using jackknife errors, dividing the field into a $3\times3$ grid and measuring the correlation function with each of the regions excluded in turn. This method is intended to give a truer estimate of the error than simple Poisson statistics. We have compared jackknife and Poisson errors and find that the two are comparable out to $\theta\approx30$\arcsec, suggesting that Poisson statistics can be used fairly reliably at small separations. At larger angular separations, where the number of pairs becomes large, Poisson statistics underpredict the error.

As a further test of our results we use simulated catalogues in place of the real sub-mm source list to measure the likelihood of our cross-correlation results arising by chance. We construct the simulated catalogues such that they show the same autocorrelation signal as the real sub-mm sources (Wei\ss\ et al., 2009).

\section{Cross-correlations with X-ray sources} \label{s-smx}

Fig.\ \ref{f-smallx} shows the result of our cross-correlation (eqn.\ 1) of 852 \emph{Chandra} X-ray sources with 126 LABOCA sub-mm sources; there is a significant positive correlation. \emph{Chandra} sources are known to be dominated by AGN at the X-ray fluxes where these sources are selected; starbursts do not make a comparable contribution until $S_{0.5-2.0}\sim10^{-17}$\xflux\ (Bauer et al., 2004), some $10\times$ fainter than the limit of the ECDFS data. Therefore, this result appears to suggest a strong AGN contribution to the sub-mm population.

Other authors have also presented cross-correlations of X-ray and sub-mm sources, as shown in Fig.\ \ref{f-smallx}. Almaini et al.\ (2003) cross-correlated \emph{Chandra} X-ray sources in the ELAIS N2 field with 17 SCUBA 850\micron\ sources and measured an extended correlation which they attributed to the two populations tracing the same large-scale structure. However, Borys et al.\ (2004) tested this result in the \emph{Hubble} Deep Field North (HDFN) and did not detect any large-scale correlation. 

Of the 19 SCUBA sources in the HDFN, Borys et al.\ (2004) report that 8 have an X-ray source within 7\arcsec; these pairs were excluded from their cross-correlation analysis, but if included would yield a positive peak in the correlation function at low separation. This is in line with our analysis, which shows a strong peak at $\theta\la10$\arcsec\ but no evidence for the strong clustering at larger separations detected by Almaini et al.\ (2003).

We find 30 X-ray sources in the ECDFS which lie within 10\arcsec\ of a sub-mm source, while a random distribution of sources would give only 7.5 such pairs (see Table \ref{t-smxray}). Assuming Poisson statistics, the detection of 30 pairs therefore represents an $8.2\sigma$ excess, and indicates that roughly a quarter of the 126 sub-mm sources have an X-ray source -- likely an AGN -- within the radius of the LABOCA beam.  In 300 simulations we found no detections of a comparable correlation, suggesting that it is unlikely this result could have been achieved by chance.

\begin{figure}
\includegraphics[width=70.mm]{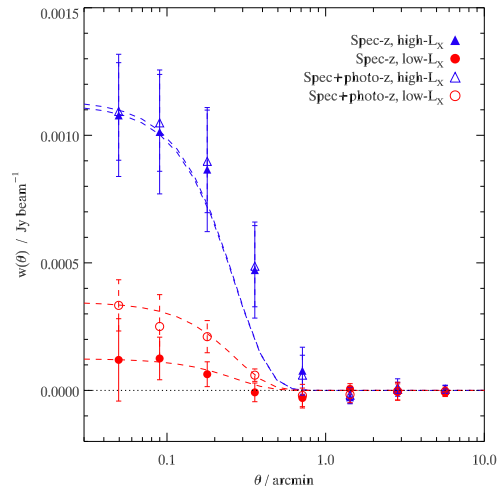}
\caption{Sub-mm stacking analysis of X-ray sources divided into high-luminosity ($L_{\rm{X}}>10^{44}$ \xlum; blue triangles) and low-luminosity ($L_{\rm{X}}<10^{44}$ \xlum; red circles). We show the results for the spec-$z$ sample only (solid), and for the sample with photo-$z$ also included (open). Sources are correlated with sub-mm pixel values to measure the excess of sub-mm flux associated with these sources. It is clear that the more X-ray luminous sources are brighter at 870\microns. We again show the LABOCA beam profile, normalised to the data; the correlations are consistent with the beam profile, suggesting the detection of source counterparts rather than associated objects. The uncertainties shown are jackknife errors.}
\label{f-smflxlum}
\end{figure}

\subsection{X-ray flux dependence} \label{ss-xflx}

In a non-unified AGN model absorbed and unabsorbed sources are intrinsically different, perhaps representing two stages of an evolutionary sequence, rather than differing purely due to orientation effects. In the absorbed phase the AGN is enshrouded by gas and dust; the unabsorbed phase occurs after this obscuring material has dissipated. Therefore, it is more absorbed AGN which are expected to be preferentially submillimetre-bright.

Models of X-ray number counts have shown that at bright fluxes unabsorbed AGN make the dominant contribution (e.g.\ Gunn, 1999; Gilli, Comastri \& Hasinger, 2007).  Based on these models, we divide the \emph{Chandra} sources at a soft-band (0.5--2 keV) flux of $S_{0.5-2.0}=1.5 \times 10^{-15}$ \xflux, yielding 685 faint X-ray sources and 167 bright sources. 

In Fig.\ \ref{f-ftbrt}\emph{a} we see that it appears certain that the faint X-ray sources show a
significant correlation with sub-mm sources while it appears that the bright
X-ray sources are less correlated. In Table \ref{t-smxray}, alongside the data for the full
X-ray sample, we also show the numbers of detected and expected X-ray-sub-mm
pairs for faint and bright sources. We confirm the strong significance of the
correlation for the faint sources. Out to $\theta<20$\arcsec\ we find 58 pairs of
sub-mm and faint X-ray sources compared to an expectation of 23.8, a $7.0\sigma$
excess assuming Poisson statistics which we have determined to be reliable at
these separations. Any excess for the bright X-ray sources is less significant,
although the statistics become poorer and the possibility that the bright and
faint X-ray sources are drawn from the same population cannot be ruled out.
Neverthelesss, it appears certain that the strong cross-correlation found in Fig.\ \ref{f-smallx} 
originates mostly in the fainter X-ray sources (see also Barger et al.,
2001).

\begin{figure*}
\includegraphics[width=170.mm]{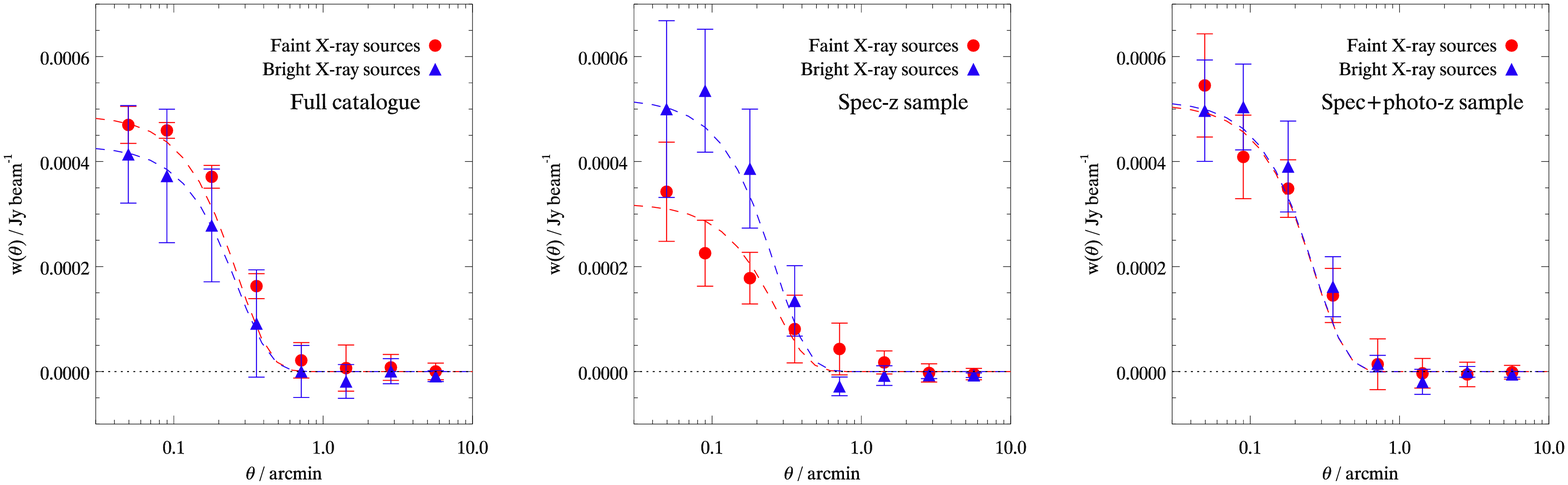}
\caption{Stacking analysis showing the average excess of sub-mm flux associated with faint ($S_{0.5-2.0}<1.5 \times 10^{-15}$ \xflux) and bright X-ray sources in the full ECDFS catalogue (\emph{a}), the spectroscopic sample (\emph{b}) and the combined spec-$z$ and photo-$z$ sample (\emph{c}). In the full catalogue, the faint sources show a higher stacked flux than the bright, but in the spec-$z$ subset the bright sources dominate in the sub-mm, suggesting this sample is not representative of the full catalogue. The combined redshift sample is better representative of the full catalogue.  We show the LABOCA beam profile (dashed/dash-dot line for faint/bright sources); the correlations are consistent with these lines, which again suggests the detection of source counterparts rather than associated objects. The uncertainties shown are jackknife errors.}
\label{f-smstacks}
\end{figure*}

Figs. \ref{f-ftbrt}\emph{b} and \emph{c} show the equivalent correlations between 
X-ray sources and the 24 and 70 micron sources in the ECDFS. Here, 
strong correlations are seen split broadly evenly between the faint and bright 
sources. If anything, the brighter X-ray sources may show a stronger correlation 
than the fainter ones, contrary to the tentative indication in the sub-mm. One 
possible explanation would be the different $k$-corrections in these wavebands -- at 870\micron\ there is a strong negative $k$-correction, increasing the detection of high-redshift sources, while at 24\microns\ and 70\microns\ the $k$-correction is positive, favouring lower redshifts. This could account for the different results in Fig.\ \ref{f-ftbrt}. To test this, we cross-correlate the 24\micron\ and 70\micron\ sources with bright and faint X-ray sources in the redshift catalogue, imposing redshift cuts of $z>0.5$ and $z>1$; we find that the results are not affected. 

Different dust temperatures might explain any difference in the sub-mm and 24--70 
micron properties of the bright and faint X-ray populations. If the faint sources include the more absorbed AGN, it may be that more absorbed sources have cooler dust temperatures -- this idea is discussed further in section \ref{s-disc}. Any temperature difference between bright and faint flux X-ray sources is unlikely to
be be driven by warmer dust in more luminous sources, since Lutz et al.\ (2010) have
shown that $L_X>10^{44}$ \xlum\ X-ray sources have brighter sub-mm fluxes. We therefore
next directly consider the luminosity dependence of the sub-mm properties of the ECDFS X-ray sources to test Lutz et al.'s result.

\subsection{$L_X$ dependence}

Fig.\ \ref{f-smflxlum} shows our sub-mm stacking analysis (eqn.\ 2) for high- and low-luminosity X-ray sources, divided at $L_X=10^{44}$ \xlum. Here we focus on a stacking analysis to be consistent with Lutz et al.\ (2010). We find a clear difference between the low and high $L_X$ sources in the sense that 
the high-$L_X$ sources are brighter in the sub-mm, confirming the result of 
Lutz et al. In this analysis we have used a catalogue of 277 X-ray sources with confirmed spectroscopic redshifts, combining sources in the catalogues of Tozzi et al.\ (2006) and Treister et al.\ (2009).

Thus  high-luminosity X-ray sources are stronger sub-mm sources than low-luminosity
X-ray sources, whereas in Fig.\ \ref{f-ftbrt} we found that fainter flux X-ray sources were 
significant sub-mm emitters. To make the comparison with Fig.\ \ref{f-smflxlum} easier, in Fig.\ \ref{f-smstacks}\emph{a} 
we show the stacked sub-mm flux around faint and bright X-ray sources. This 
again shows that the faint-flux sources are at least as powerful sub-mm sources as 
the bright. 

However, this comparison between the flux and luminosity results could also be 
affected by the incompleteness of the spectroscopic redshift (spec-$z$) sample used to
produce the $L_X$ result. In Figs.\ \ref{f-smstacks}\emph{b} and \emph{c} we show the equivalent results for the spec-$z$ and 
photometric samples. Comparing Fig.\ \ref{f-smstacks}\emph{b} to Fig.\ \ref{f-smstacks}\emph{a}, it seems that the spec-$z$ sample 
looks more similar to the $L_X$ result in Fig.\ \ref{f-smflxlum}, with the brighter flux sources appearing 
more sub-mm bright than their fainter counterparts. Therefore, the spec-$z$ sample may not be 
a representative subset of the X-ray sources in terms of the sources' sub-mm properties.

Comparing the left and central panels of Fig.\ \ref{f-smstacks} suggests that there is a population of \emph{Chandra} sources in the full ECDFS catalogue which have faint X-ray fluxes but are bright at 870\micron, which are missing from the spec-$z$ sample. If these faint sources also have lower X-ray luminosities, it would cast some doubt on the idea that sub-mm emission arises from high-luminosity sources only. We therefore next consider the differences between the spectroscopic sample and the full catalogue.

\subsection{Incompleteness of the spectroscopic sample}

\begin{figure}
\includegraphics[width=70.mm]{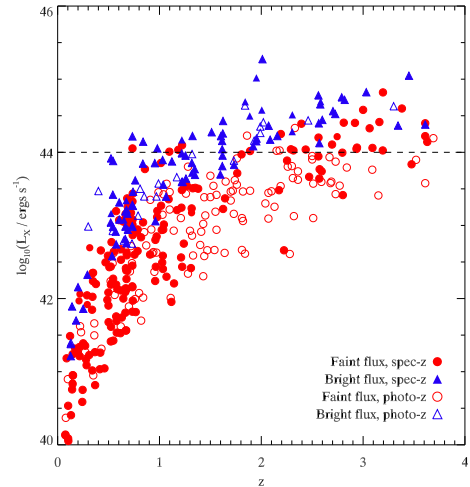}
\caption{X-ray luminosity as a function of redshift for ECDFS sources in the spec-$z$ sample (filled points) and photo-$z$ sample (open points). Faint X-ray sources are shown as red circles, bright X-ray sources as blue triangles. The line at $L_{\rm{X}}=10^{44}$ \xlum\ shows the luminosity limit which has been found to separate submm-bright and submm-faint sources. There is a notable defincency of faint-flux spec-$z$ sources in the range $1.3<z<2.5$.}
\label{f-lumz}
\end{figure}

We have seen that there is some indication of a difference between the sub-mm properties of X-ray sources in the full catalogue and of those in the spec-$z$ subset, and we therefore attempt to identify ways in which the latter is unrepresentative of the former. We first find that the spec-$z$ sample is missing faint-flux, low-luminosity X-ray sources at high redshifts, with a particularly significant dearth of faint X-ray sources in the range $1.3<z<2.5$ (Fig.\ \ref{f-lumz}).

In this redshift range the spec-$z$ sample consists almost entirely of bright X-ray sources, whereas the full sample would be expected to also include faint sources here. Submillimetre-bright objects are known to lie at high-redshift, with an $n(z)$ peaking at $z\approx2$ (Chapman et al., 2005), so the deficiency of faint-flux, low-luminosity sources in this redshift range in the spec-$z$ sample could be significant.

\begin{figure}
\includegraphics[width=70.mm]{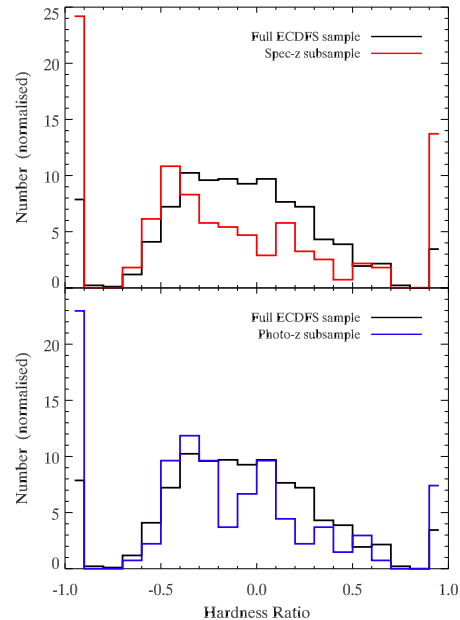}
\caption{The hardness ratio distribution of sources in the full ECDFS X-ray catalogue compared to those in the spec-$z$ sample (top panel) and photo-$z$ sample (bottom panel), all normalised to the same total number of sources. The spec-$z$ subset significantly undersamples the harder sources in the ECDFS, while the photo-$z$ sample contains relatively more hard sources and is better representative.}
\label{f-hardness}
\end{figure}

The upper panel of Fig.\ \ref{f-hardness} highlights another aspect of the difference between the full and spec-$z$ samples. There is an appreciable difference in the hardness of the sources in the full catalogue and the spec-$z$ subset, with the latter significantly undersampling the harder sources. A Kolmogorov--Smirnov (KS) test indicates a 0.26\% likelihood that the two distributions are drawn from the same population, falling to 0.1\% if the spikes at $HR=\pm1$ are excluded from the test. Since the difference between Figs.\ \ref{f-smstacks}\emph{a} and {b} may indicate that there exists a population of faint X-ray, submillimetre-bright sources without measured spectroscopic redshifts, the indication found here that the sources missing from the spec-$z$ sample may typically be high-redshift, absorbed AGN provides an interesting suggestion that such sources are bright at 870\microns.

We previously limited our redshift catalogue to only include secure spectroscopic redshifts measured by Tozzi et al.\ (2006) and Treister et al.\ (2009). We now add to the 277 spec-$z$ sources 114 sources with photometric redshifts, also presented in Tozzi et al.'s catalogue. The distribution of hardness ratios for the photo-$z$ sources is shown in the lower panel of Fig.\ \ref{f-hardness} and it is clear that they better represent the overall distribution than the spec-$z$ sources do. When added to the spec-$z$ sample to give a combined redshift catalogue of 391 objects, a KS test yields a 2.3\% likelihood that the redshift catalogue and the full catalogue sample the same population. Though still small, this probability is an order of magnitude greater than found previously. Additionally, Fig.\ \ref{f-lumz} shows that the combined redshift catalogue does not exhibit the same deficiency of faint-flux, low-luminosity sources in the $1.3<z<2.5$ range. In the right panel of Fig.\ \ref{f-smstacks}, we show that, having included these extra photo-$z$ sources, the sub-mm stacking analysis for faint-flux objects now better resembles that found with the full sample.

However, while the sub-mm contribution from faint-flux X-ray sources has increased, Fig.\ \ref{f-smflxlum} shows that this does not yield a substantial increase in the contribution from the X-ray sources with low intrinsic luminosities, which remains significantly below that of the high-luminosity sources. This suggests that the reduced sub-mm contribution found for faint-flux sources in the spec-$z$ sample (Fig.\ \ref{f-smstacks}\emph{b}) was not simply the result of undersampling the low-luminosity X-ray sources. Instead, we continue to find that high-luminosity sources dominate the AGN sub-mm contribution. 

\begin{figure*}
\includegraphics[width=140.mm]{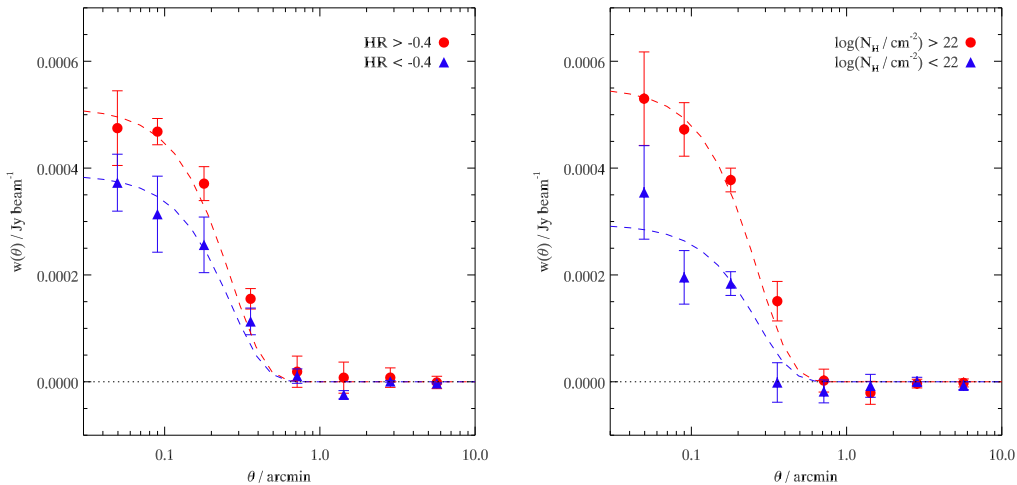}
\caption{Sub-mm stacking analyses for ECDFS X-ray sources. (\emph{a}) X-ray sources are divided into hard (red circles) and soft (blue triangles) sources at a hardness ratio of $-0.4$. (\emph{b}) X-ray sources are divided into absorbed (red circles) and unabsorbed (blue triangles) at a column density of $N_H=10^{22}$ cm$^{-2}$; this is done by combining hardness ratio and redshift information, as described in the text. Jackknife errors are shown.}
\label{f-xcorrhrnh}
\end{figure*}

This means that many of these high-luminosity sources must also be faint in X-ray flux. There is of course no necessary contradiction here, since these sources are expected to lie at high redshift. If there is also a preference for more absorbed AGN to be sub-mm bright, this would compound the effect, with absorption making the high-luminosity sources appear fainter still in X-ray flux. Of course, the full redshift sample still has less than half the total number of X-ray sources, so it remains conceivable that the sub-mm contribution of the low-$L_X$ population could rise if these were to be included in Fig.\ \ref{f-smflxlum}.

\subsection{Hardness ratio dependence} \label{ss-hrat}

We now make a more direct test of the hypothesis that absorbed X-ray sources are strong sub-mm emitters by performing the sub-mm stacking analysis and cross-correlation for X-ray sources divided by hardness ratio. The stacking analysis is shown in Fig.\ \ref{f-xcorrhrnh}\emph{a} and although the significance is marginal, if anything it is the harder sources which contribute more than the softer ones. Our standard cross-correlation method (eqn. 1) gives a similar result.

The division between `harder' and `softer' sources in Fig.\ \ref{f-xcorrhrnh}\emph{a} is at $HR=-0.4$, which corresponds to a hydrogen column density of $N_H=10^{22}$ cm$^{-2}$ at $z=1$, assuming a power-law spectrum with $\Gamma=-2$. This column density is an appropriate place to divide the sample, since a non-unified AGN model (see \S\ref{s-ebl}) predicts that $N_H>10^{22}$ cm$^{-2}$ sources are typically more sub-mm bright, while in the unified AGN model we should find that $N_H<10^{22}$ cm$^{-2}$ and $N_H>10^{22}$ cm$^{-2}$ sources have the same average sub-mm flux.\footnote{This fiducial column of $N_H=10^{22}$ cm$^{-2}$ is also consistent with the result of Page et al.\ (2004), who find that quasars more absorbed than this limit dominate the sub-mm contribution.}

\begin{figure}
\includegraphics[width=70.mm]{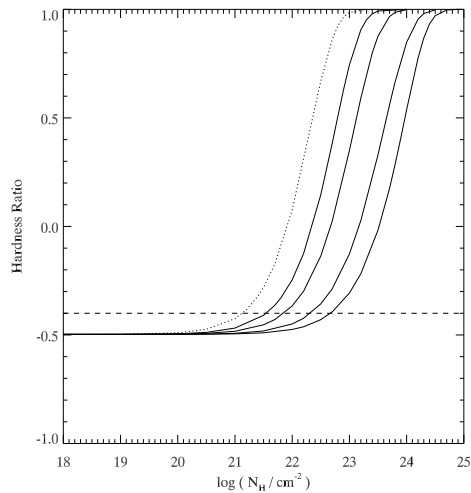}
\caption{Hardness ratio vs.\ hydrogen column density for AGN at different redshifts, generated with the \emph{Chandra} PIMMS tools assuming a power-law spectrum with $\Gamma=-2$. The dotted line on the left represents $z=0$; moving right, the next four lines are for $z=0.5$, 1.0, 2.0 and 3.0. The dashed line at $HR=-0.4$ shows where we divide the sample for the analysis in Fig.\ \ref{f-xcorrhrnh}\emph{a}.}
\label{f-hr-nh}
\end{figure}

Therefore, the finding that the stacked fluxes of the harder and softer sources are similar in Fig.\ \ref{f-xcorrhrnh}\emph{a} appears to be supportive of the unified AGN model. However, this analysis does not take into account the effect of redshift on hardness ratio. As Fig.\ \ref{f-hr-nh} shows, at $z\ge2$ the $HR<-0.4$ X-ray sources do include very absorbed ($\log(N_H / \rm{cm}^{-2})=22$--23) sources. It is therefore not certain that the sub-mm correlation measured for  $HR<-0.4$ X-ray sources in Fig.\ \ref{f-xcorrhrnh}\emph{a} arises from the low-column, less absorbed population; it could instead be due to highly absorbed AGN at high-redshift. To break the degeneracy between redshift and column we must use the X-ray sources with known redshifts.

We divide the X-ray redshift sample (using both spec-$z$ and photo-$z$) into four bins: $z=0.25$--0.75, 0.75--1.5, 1.5--2.5 and 2.5--3.5. In each bin sources are split into hard and soft, with the divisions made at $HR=-0.25$, $-0.37$, $-0.45$ and $-0.47$, these being the hardness ratios corresponding to $N_H=10^{22}$ cm$^{-2}$ for $z=0.5$, 1.0, 2.0 and 3.0, respectively (Fig.\ \ref{f-hr-nh}). This gives us a sample of $N_H<10^{22}$ cm$^{-2}$ sources and $N_H>10^{22}$ cm$^{-2}$ sources. 

The sub-mm stacking analyses for these absorbed and unabsorbed samples are shown in Fig.\ \ref{f-xcorrhrnh}\emph{b}. There is again an indication here that the more absorbed sources are typically brighter in the sub-mm. Comparing the two panels of  Fig.\ \ref{f-xcorrhrnh}, it appears that some of the sub-mm flux associated with the softer X-ray sources arises from the absorbed, $N_H>10^{22}$ cm$^{-2}$ population. 

Lutz et al.\ (2010) also investigated how sub-mm flux depends on column density for ECDFS X-ray sources and found average stacked fluxes for the  $N_H<10^{22}$ cm$^{-2}$ and $N_H>10^{22}$ cm$^{-2}$ populations which are in good agreement our measurements in Fig.\ \ref{f-xcorrhrnh}\emph{b}. 

In summary, then, our results have shown marginal indications that fainter ($S_{0.5-2.0}<1.5\times10^{-15}$ \xflux) or harder ($HR>-0.4$) X-ray sources correlate more strongly with the sub-mm, and a stronger suggestion that more absorbed ($N_H>10^{22}$ cm$^{-2}$) X-ray sources are preferentially sub-mm bright. The strongest dependence appears to be on X-ray luminosity, with $L_X>10^{44}$ \xlum\ sources being significantly brighter in the sub-mm. All of these results will be compared to AGN model predictions in \S\ref{s-ebl}.

\begin{figure}
\includegraphics[width=70.mm]{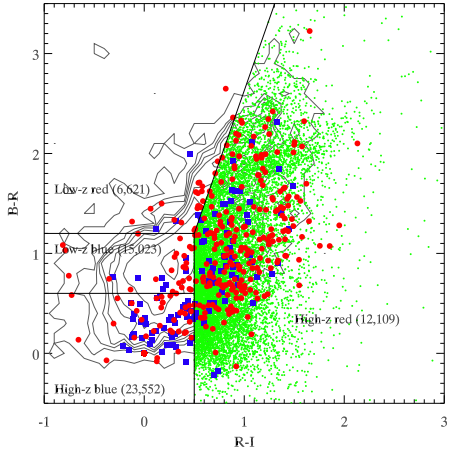}
\caption{$BRI$ colour-colour plot for sources in the ECDFS.The contours show the colour distribution of all optical sources in the ECDFS MUSYC catalogue. The regions marked by the solid black lines indicate where we selected high-$z$ blue, low-$z$ blue, low-$z$ red and high-$z$ red galaxies, as labelled. The number of objects in each group is also given. Faint and bright X-ray sources are marked as red circles and blue squares, respectively, and green points are objects selected as high-$z$ red galaxies. }
\label{f-brixray}
\end{figure}

\section{Cross-correlation with optical populations} \label{s-optcorr}

We next cross-correlate ECDFS sub-mm sources with optical sources from the MUSYC $BVR$-selected catalogue. We divide the optical sources into four populations (Fig.\ \ref{f-brixray}) according to their $BRI$ colours, intended to correspond to low-$z$ red, high-$z$ red, low-$z$ blue and high-$z$ blue galaxies, where high-$z$ refers to $z>0.5$ galaxies. The colour criteria used to divide the four sections were defined using well established galaxy evolution models (Metcalfe et al., 2001, 2006).

\begin{figure*}
\includegraphics[width=170.mm]{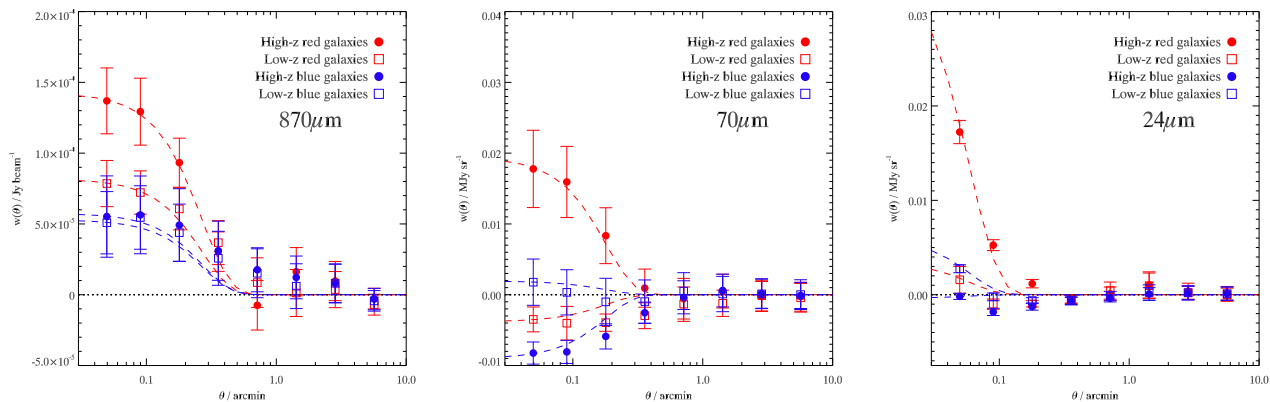}
\caption{Stacking analyses for our four optically-defined galaxy populations at 870\microns\ (\emph{a}), 70\microns\ (\emph{b}) and 24\microns\ (\emph{c}). The high-$z$ red galaxies have the brightest stacked flux in all three wavebands. The units of the vertical axes are the same as in the respective intensity maps: Jy beam$^{-1}$ at 870\microns\ and MJy sr$^{-1}$ in the other bands. We convert to $\umu$Jy in Table \ref{t-seds}. For each correlation we show the beam profile of the detector, approximated by a Gaussian with $\rm{FWHM}=27$\arcsec, 18\arcsec\ and 6\arcsec\ for 870\micron, 70\micron\ and 24\micron\ respectively.}
\label{f-optstacks}
\end{figure*}

We find that of these four galaxy populations, the high-$z$ red galaxies cross-correlate most strongly with 870\micron\ sources and also have the brightest stacked sub-mm flux. The left panel of Fig.\ \ref{f-optstacks} shows that the high-$z$ red galaxies are strongly detected at 870\micron\ at $5.9\sigma$ significance with $S_{870}=142\pm24 \umu$Jy. The low-$z$ red galaxies also show a significant detection at $4.7\sigma$ significance while the blue galaxy populations are more weakly detected at $\sim2\sigma$ significance (see Table \ref{t-seds}).

Fig.\ \ref{f-optstacks} and Table \ref{t-seds} also show the stacking analyses at 70\microns\ and 24\microns. In both of these wavebands, the high-$z$ red galaxies are the only sources to show any significant flux. Taken together, the analyses shown in Fig.\ \ref{f-optstacks} suggest that the $z>0.5$ red galaxies emit most strongly at long IR wavelengths. 

\begin{table}
\caption{Average 870\micron, 70\micron\ and 24\micron\ fluxes of different X-ray groups and optical galaxy populations, found by stacking the images. Positive detections at $>3\sigma$ significance are indicated in bold.}
\label{t-seds}
\begin{tabular}{|c|c|c|c|}
\hline
Population & 870\micron\ & 70\micron\ & 24\micron\ \\
& $\umu$Jy & $\umu$Jy & $\umu$Jy    \\
\hline
All X-ray sources       & $\mathbf{477\pm28}$ & $\mathbf{758\pm132}$ & $\mathbf{126\pm11}$ \\
Faint X-ray sources   & $\mathbf{488\pm37}$ & $\mathbf{759\pm115}$ & $\mathbf{120\pm14}$ \\
Bright X-ray sources & $\mathbf{430\pm97}$ & $\mathbf{754\pm107}$ & $\mathbf{212\pm30}$ \\
\hline
High-$z$ red galaxies &   $\mathbf{142\pm24}$ & $\mathbf{111\pm28}$ & $\mathbf{23\pm3}$ \\
Low-$z$ red galaxies &    $\mathbf{81\pm17}$ & $-22\pm14$ & $2\pm2$ \\
High-$z$ blue galaxies & $57\pm30$ & $-51\pm13$ & $0\pm0$ \\
Low-$z$ blue galaxies &  $53\pm23$ & $11\pm22$   & $\mathbf{4\pm1}$ \\
\hline
\end{tabular}
\end{table}

Almaini et al.\ (2005) suggested that the explanation for a cross-correlation between bright sub-mm sources and $R<23$, $z\approx0.5$ galaxies could be gravitational lensing. But where Almaini et al.\ found a cross-correlation between $S_{850}>10$ mJy sub-mm sources and galaxies and little cross-correlation for fainter sub-mm sources, we find that the cross-correlation with high-$z$ red galaxies is dominated by $S_{850}<10$ mJy sub-mm sources, so any such lensing signature is absent from our data.

We have compared the optical (and NIR -- see Hill, 2010) colours of the high-$z$ red galaxies to those of the X-ray sources and find that the faint X-ray sources and the high-$z$ red galaxies tend to occupy the same colour space, while the locus of the bright X-ray sources is elsewhere (Fig.\ \ref{f-brixray}). This raises the possibility that the high-$z$ red galaxies and the faint X-ray sources, the two object classes seen to exhibit the strongest correlation with sub-mm sources, both sample the same population. The faint X-ray sources are expected to include the more absorbed AGN; the question, then, is whether the high-$z$ red galaxies also host absorbed AGN, specifically Compton-thick AGN which are X-ray-undetected.

We have attempted to estimate the AGN fraction within the high-$z$ red population using the \emph{Spitzer} IRAC [3.6]--[4.5]:[5.8]--[8.0] colour-colour plot, which is known to be a robust way of distinguishing AGN from other sources (Stern et al., 2005). However, IRAC observations are available only for the small, central field (SWIRE survey; Lonsdale et al., 2004) and the 8\micron\ resolution is relatively poor, so only $\sim1$\% of our X-ray and optical populations are detected in all four IRAC bands. 70\% of the faint X-ray sources and 3\% of the high-$z$ red galaxies which are detected in all four IRAC channels lie within or very close to the AGN wedge, but with such small samples it is hard to draw any firm conclusions here.

\begin{figure*}
\includegraphics[width=125.mm]{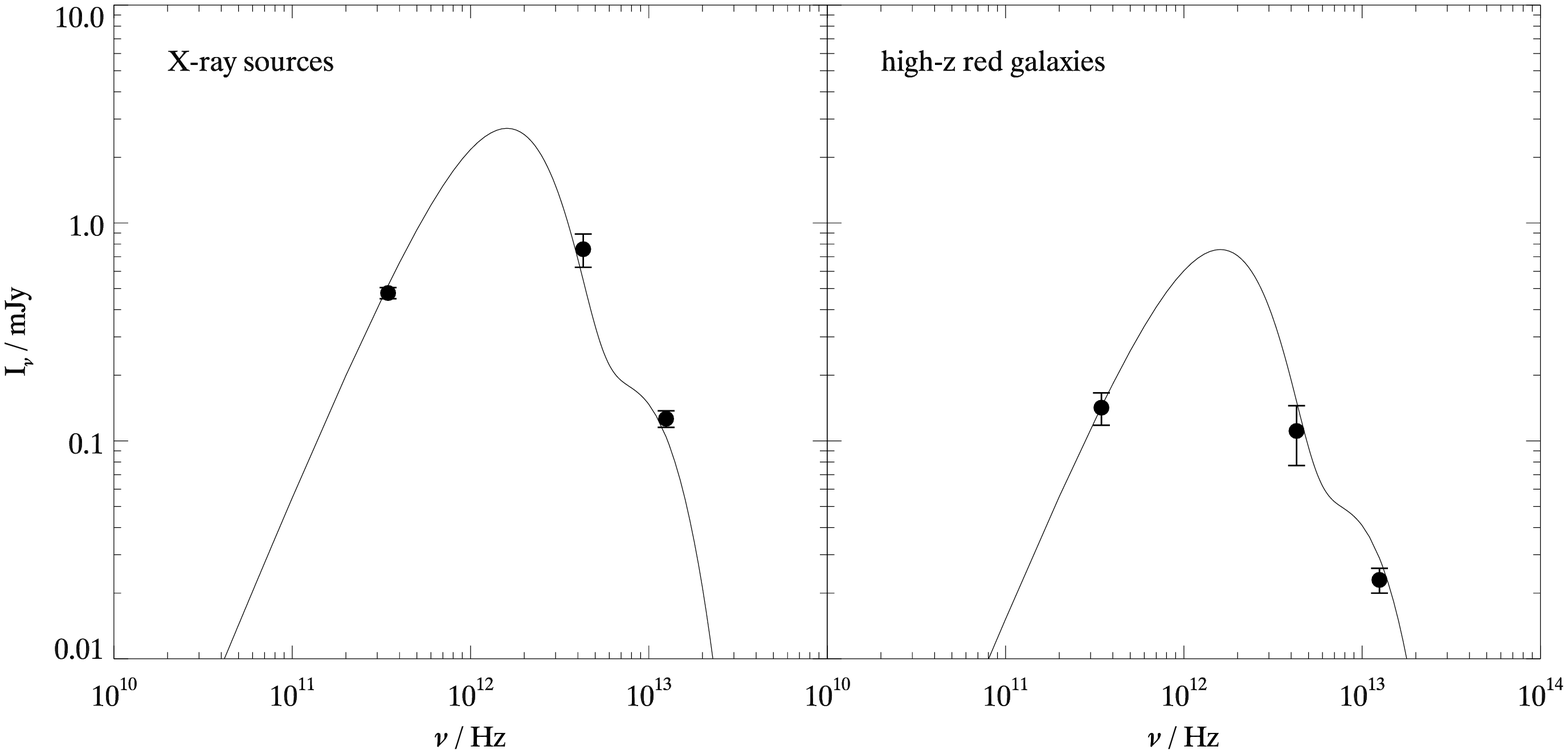}
\caption{Stacked 870\micron, 70\micron\ and 24\micron\ fluxes for (\emph{a}) X-ray sources and (\emph{b}) high-$z$ red galaxies in the ECDFS. In both cases the data are well fit by a two-blackbody dust model with $T_1=27$ K, $T_2=120$ K and a ratio between the two components of $L_2/L_1=0.3$. With only 3 points in each case these temperatures cannot be taken too literally, however it is clear there is no inconsistency here between the FIR colours of the two populations.}
\label{f-seds}
\end{figure*}

Looking at longer, mid/far-IR wavelengths, we can use the stacked 24\micron, 70\micron\ and 870\micron\ fluxes given in Table \ref{t-seds} to characterise the average far-IR SEDs of the high-$z$ red galaxies and the X-ray sources (Fig.\ \ref{f-seds}). We find that two temperature components are needed to fit the data for the X-ray sources and the high-$z$ red galxies galaxies, with both populations well matched by a model with $T_1=27$ K, $T_2=120$ K and a ratio between the components of $L_2/L_1=0.3$, although this is by no means a unique solution. With only three datapoints constraining the fits the derived temperatures should not be taken too seriously, particularly as the 24\micron, 70\micron\ and 870\micron\ fluxes may yet arise from different populations. Nevertheless, Fig.\ \ref{f-seds} does serve to illustrate that the FIR colours (and implied dust temperatures) of red galaxies and X-ray sources appear broadly consistent.

We conclude that there is no inconsistency in the optical/NIR/FIR colours of $z>0.5$ red galaxies and faint X-ray sources. These galaxies could therefore host Compton-thick AGN. We shall see later that the sub-mm background associated with these red galaxies is also consistent with the predicted contribution from such AGN. 

\section{Submillimetre number counts and extragalactic background} \label{s-ebl}

Using the sub-mm stacking analyses presented for X-ray sources and high-$z$ red galaxies, we can quantify the contribution made by these populations to the sub-mm background, by multiplying by the stacked average flux by the total number of sources and dividing by 0.25 deg$^2$. For the high-$z$ red galaxies this contribution is $4.8 \pm 1.2$ Jy deg$^{-2}$, while for the X-ray sources it is $1.5 \pm 0.1$ Jy deg$^{-2}$.

The total sub-mm background, based on \emph{COBE FIRAS} measurements, is estimated to be $\sim45$ Jy deg$^{-2}$ (Puget et al., 1996; Fixsen et al., 1998). Therefore, the value of $1.5 \pm 0.1$ Jy deg$^{-2}$ from X-ray sources represents only a very small fraction of the total extragalactic background light (EBL). However, the total contribution that could be made by active galactic nuclei, a significant proportion of which could be missing from the X-ray sample we have used, could yet be far greater. In order to quantify this, we make use of a model of obscured AGN.

\subsection{Obscured AGN model}

We have detected significant sub-mm flux from X-ray sources which are likely to be a population of obscured AGN. We now wish to compare the amount of detected sub-mm flux to the predictions of an obscured AGN model which also fits the X-ray background (XRB). The model will also allow us to estimate the amount of sub-mm flux which is predicted to arise in Compton-thick AGN which will not be detected as X-ray sources but whose hosts may be detected as individual galaxies, for instance the $z>0.5$ red galaxy population we have also found to be a significant contributor to the sub-mm background.

The model we employ is that of Gunn \& Shanks (1999), which is known to match the observed hard X-ray background. It assumes that AGN are drawn from an intrinsically flat distribution of seven absorbing columns between $N_{\rm{H}}=10^{19.5}$ cm$^{-2}$ (essentially unobscured) and $10^{25.5}$ cm$^{-2}$ (heavily Compton-thick). 

The model as presented by Gunn \& Shanks (1999; see also Gunn, 1999) is non-unified: it assumes that the observed column density is directly a measure of the intrinsic amount of gas, not a result of viewing angle. This assumption has no bearing on the model's success in fitting the XRB, but is significant in the sub-mm since dust masses in the model are calculated from a gas-to-dust ratio, so more heavily absorbed sources will have greater dust masses and therefore make a greater contribution in the sub-mm. We make use of the model within this non-unified paradigm, but also later adapt it in order to explore the unified AGN scenario, in which different column densities arise due to orientation effects and do not reflect any intrinsic difference between AGN populations. 

In both cases we assume a dust temperature of 30K, consistent with observations of sub-mm sources (Coppin et al., 2008b; Elbaz et al., 2010) and an emissivity index of $\beta=1.5$ (Dunne \& Eales, 2001).

In Fig.\ \ref{f-850cnts} we show the prediction of the non-unified model, which gives an excellent fit to the sub-mm source counts at $S_{850}>0.5$ mJy. At fainter 850\micron\ fluxes, the AGN model begins to underpredict the counts, however we show in the figure the predicted contribution of normal galaxies, which is significant at these faint fluxes (see also Busswell \& Shanks, 2001). The predicted density of bright sub-mm galaxies (SMGs; $S_{850}>4$ mJy) is 783 deg$^{-2}$, in very good agreement with the error-weighted mean of $768\pm89$ deg$^{-2}$ for the observational data shown in Fig.\ \ref{f-850cnts}.\footnote{This error-weighted mean is calculated without the ECDFS counts (Wei\ss\ et al., 2009) as this field is known to contain an underdensity of bright sub-mm sources.}

\subsection{Model predictions for ECDFS X-ray sources}

\subsubsection{Sub-mm contribution vs.\ X-ray flux}

The $1.5 \pm 0.1$ Jy deg$^{-2}$ sub-mm background contribution made by the ECDFS X-ray sources provides a crucial observational constraint on our obscured AGN model, as we can use the model to generate a predicted sub-mm background contribution from only those AGN with 0.5--2.0 keV X-ray fluxes greater than the $1.1\times 10^{-16}$ \xflux\ limit of the ECDFS survey. 

\begin{figure*}
\includegraphics[width=120.mm]{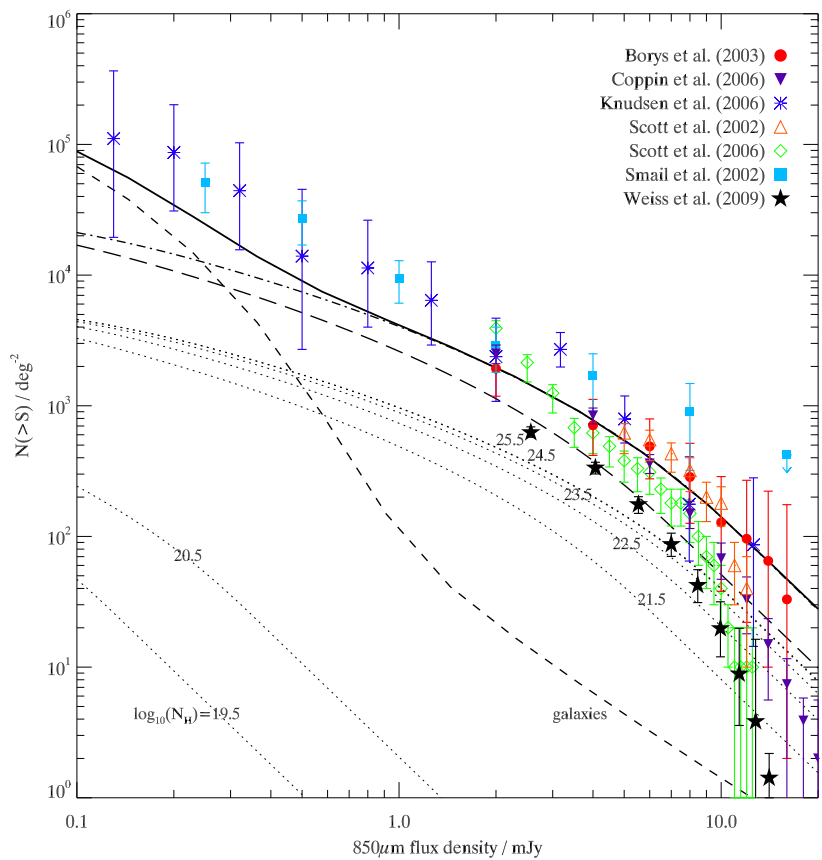}
\caption{Integral number counts at 850\microns. We use the model of Gunn \& Shanks (1999), which is known to fit the hard X-ray background, to generate predicted sub-mm number counts from a population of obscured AGN. The total AGN contribution is shown by the dash-dot line; dotted lines show individual contributions from AGN with different columns (labelled; note that the $N_{\rm{H}}=10^{24.5}$ and $10^{25.5}$ cm$^{-2}$ lines are indistinguishable); this is a non-unified AGN model where more absorbed AGN contribute more strongly to the sub-mm. The short-dashed line shows the predicted contribution from galaxies and the solid line is the total prediction, i.e. the sum of the galaxy and AGN contributions. Finally, the long-dashed line shows the alternative total AGN prediction, using the model normalised to the observed sub-mm background from ECDFS X-ray sources.}
\label{f-850cnts}
\end{figure*}

We find that using the model shown in Fig.\ \ref{f-850cnts}, this predicted background is 2.3 Jy deg$^{-2}$, $\approx50$\% greater than the measured value. This initially appears to indicate that our  model is over-predicting the contribution which AGN make in the sub-mm. However, the bright sub-mm source counts in the ECDFS are known to be lower than in other fields, by as much as a factor of 2 (Wei\ss\ et al., 2009). This deficiency of bright sub-mm sources in the ECDFS may explain why the total stacked sub-mm flux of ECDFS AGN is lower than the model prediction.

Nevertheless, we can revise the model to bring the predicted sub-mm background from X-ray-detected AGN into line with the observations. We make this revision by altering the dust covering factor, which is the only remaining free parameter in the model after fixing the dust temperature at 30K and the emissivity index at $\beta=1.5$. Gunn \& Shanks assume a covering factor $f_{cov}=1.0$, which represents an isotropic distribution of dust. If we instead take $f_{cov}=0.65$, the predicted sub-mm background from X-ray-detected AGN falls to 1.5 Jy deg$^{-2}$, consistent with the observational value.

In this revised model, however, the total AGN contribution to the sub-mm sources counts is lower (long-dashed line in Fig.\ \ref{f-850cnts}). The model now accounts for only around half of bright sources: the predicted number of SMGs falls from the 783 deg$^{-2}$ quoted above to 367 deg$^{-2}$. However, while this value is lower than most of the observational data, it agrees very well with the observed SMG sky density of $335\pm34$ deg$^{-2}$ for the ECDFS. 

This is significant, because it means that our non-unified AGN model simultaneously matches the total ECDFS  870\micron\ number counts and the measured sub-mm background flux from the ECDFS X-ray sources. Therefore, our original model, which matches the total sub-mm source counts well and predicts a background contribution of 2.3 Jy deg$^{-2}$ from $S_{0.5-2.0}\ge1.1\times 10^{-16}$ \xflux\ sources, may represent a more valid description of the AGN contribution to the sub-mm population in other more typical fields.

\begin{figure}
\includegraphics[width=70.mm]{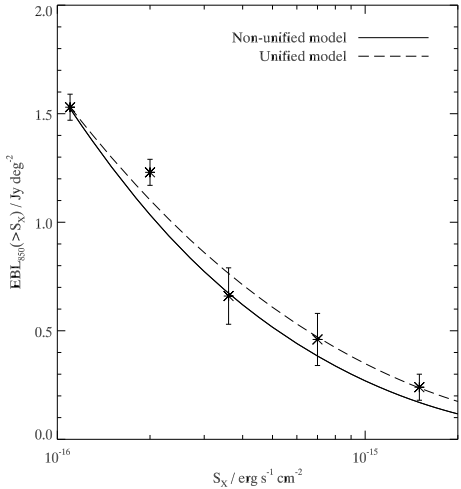}
\caption{The contribution of AGN to the sub-mm extragalactic background light (EBL) as a function of limiting AGN X-ray flux in the 0.5--2.0 keV band. Predictions from the obscured AGN model are shown for both the unified and non-unified cases. Datapoints show observational measurements from the ECDFS data.}
\label{f-eblflux}
\end{figure}

Taking the model which matches the measured $1.5 \pm 0.1$ Jy deg$^{-2}$ sub-mm flux from  $S_{0.5-2.0}\ge1.1\times 10^{-16}$ \xflux\ X-ray sources, we can make further tests by predicting the expected sub-mm background contribution from X-ray sources in other X-ray flux ranges. This is shown in Fig.\ \ref{f-eblflux}. We measure the actual sub-mm background associated with ECDFS sources to several X-ray flux limits and compare to the predictions. The figure indicates that the model is predicting the trend with flux generally very well. Since the data in Fig.\ \ref{f-eblflux} include those in Fig.\ \ref{f-smstacks}\emph{a}, this also shows that our AGN model fits the data in that figure.

\subsubsection{Sub-mm contribution vs.\ $L_X$}

We can also test the model's predictions against the result in Fig.\ \ref{f-smflxlum}, where we found that $L_X>10^{44}$ \xlum\ sources are brighter in the sub-mm. The AGN model naturally predicts such a luminosity dependence, since the amount of light that can be absorbed in the X-ray/optical (and subsequently reradiated in the sub-mm) is proportional to the intrinsic X-ray/optical luminosity. A simple test we can do with the model is to calculate the expected sub-mm background from AGN which are brighter or fainter than the characteristic luminosity $L^{*}$. This prediction should be consistent with the result from Fig.\ \ref{f-smflxlum} since $L^*\approx10^{44}$ \xlum\ at the $z\approx2$ redshift expected for sub-mm sources (Aird et al., 2010). Our model predicts a sub-mm background of 0.6 Jy deg$^{-2}$ from $L>L^*$ ECDFS X-ray sources and 0.9 Jy deg$^{-2}$ from $L<L^*$ sources (although high-luminosity sources are much brighter in the sub-mm, there are many more low-luminosity sources). Taking the stacked fluxes in Fig.\ \ref{f-smflxlum} and scaling to account for the incompleteness of the redshift sample, we measure a sub-mm background contribution from $L_X>10^{44}$ \xlum\ sources of $0.6\pm0.1$ Jy deg$^{-2}$ and for lower-luminosity sources of $0.9\pm0.3$ Jy deg$^{-2}$, in excellent agreement with the predictions.

\subsubsection{Sub-mm contribution vs.\ $N_H$}

Finally, we test the model against the result in Fig.\ \ref{f-xcorrhrnh}\emph{b}, where we determined stacked fluxes for X-ray sources more or less absorbed than $N_H=10^{22}$ cm$^{-2}$. The model predicts sub-mm background contributions of 1.0 Jy deg$^{-2}$ and 0.5 Jy deg$^{-2}$ for ECDFS X-ray sources with $N_H>10^{22}$ cm$^{-2}$ and $N_H<10^{22}$ cm$^{-2}$ respectively. These agree well with the measured values of $1.2\pm0.2$ Jy deg$^{-2}$ and $0.3\pm0.1$ Jy deg$^{-2}$ respectively. We summarise all of these tests in Table \ref{t-tests}, where we also compare to predictions from the unified AGN model, which is described next.

\begin{table}
\caption{Contributions to the sub-mm extragalactic background from ECDFS X-ray sources, divided by X-ray flux, X-ray luminosity and hydrogen column density. In each case we show the measured value for our stacking analyses, followed by the predictions of our AGN model in both the non-unified and unified cases. All sub-mm background contributions are given in Jy deg$^{-2}$; the X-ray fluxes are soft band ($0.5-2.0$ keV) fluxes in units of \xflux.}
\label{t-tests}
\begin{tabular}{|c|c|c|c|}
\hline
& & \multicolumn{2}{c}{Model prediction} \\
Population & Observation & \emph{Non-unified} & \emph{Unified} \\
\hline
$S>1.5\times10^{-15}$ & $0.2\pm0.1$ & $0.1$ & $0.2$ \\
$S<1.5\times10^{-15}$ & $1.3\pm0.1$ & $1.4$ & $1.3$ \\
\hline
$L_X>L^*$ & $0.6\pm0.1$ & $0.6$ & $0.8$ \\
$L_X<L^*$ & $0.9\pm0.3$ & $0.9$ & $0.7$ \\
\hline
$N_H>10^{22}$ cm$^{-2}$ & $1.2\pm0.2$ & $1.0$ & $0.4$ \\
$N_H<10^{22}$ cm$^{-2}$ & $0.3\pm0.1$ & $0.5$ & $1.1$ \\
\hline
\end{tabular}
\end{table} 

\subsection{Unified AGN model} \label{ssec-unif}

In the unified case, we retain the same distribution of columns as before, since this is still important when considering the X-ray fluxes of the AGN, however the column density is no longer an indicator of the total absorbed luminosity and therefore no longer affects the sub-mm flux.

Having applied the observational constraint that the X-ray-detected AGN should contribute only 1.5 Jy deg$^{-2}$ to the sub-mm background, this unified model underpredicts the sub-mm number counts by as much as an order of magnitude: the total predicted number of $S_{850}>4$ mJy sources is 72 deg$^{-2}$, $\approx5$ times lower than the ECDFS value and $\approx10$ times lower than the average of the other fields.

The significance of AGN to the sub-mm population is therefore considerably less within the framework of the unified AGN model than in the non-unified case. While we showed that in the non-unified model obscured AGN can make a dominant contribution, potentially even fully matching the source counts, in the unified model AGN account for only 10--20\% of bright sub-mm sources.

We make the same tests of this model which were described previously for the non-unified case. The predictions are shown in Table \ref{t-tests} and Fig.\ \ref{f-eblflux} alongside those for the non-unified model and the observational measurements. We note that the predictions of the unified model for the flux and luminosity samples are in generally good agreement with the data. In these cases, the difference between the unified and non-unified cases are not great enough to say that the observations prefer one to the other. As expected, though, the unified and non-unified cases show a more significant difference in the predicted contributions from more and less absorbed X-ray sources, and in this case it is the non-unified model which appears to better fit the observations.

\subsection{Submillimetre background predictions}

Having constrained the models based on the sub-mm background arising from X-ray-detected AGN, we can generate a prediction for the total contribution from all AGN. Down to a limit of $S_{850 }=1$ mJy, the total measured sub-mm background in the ECDFS LABOCA data is 13.6 Jy deg$^{-2}$ (based on the $\alpha=-3.2$ power law determined from the $P(D)$ analysis presented by Wei\ss\ et al., 2009). 

To this flux limit, the non-unified model with $f_{cov}=0.65$ predicts a total contribution to the EBL of 6.8 Jy deg$^{-2}$, 50\% of the measured value. The original non-unified model with $f_{cov}=1.0$ predicts a background contribution of 12.2 Jy deg$^{-2}$, $\approx90$\% of the total EBL. 

At very faint 850\micron\ fluxes, spiral galaxies may make a dominant contribution to the number counts, so we find that the fraction of the sub-mm EBL contributed by AGN decreases at fainter flux limits. The total sub-mm background, from observations with \emph{COBE}, is $\approx45$ Jy deg$^{-2}$. According to our model, the total sub-mm background from AGN is predicted to be 11.2 Jy deg$^{-2}$ in the case with $f_{cov}=0.65$, or 17.7 Jy deg$^{-2}$ with $f_{cov}=1.0$. The former value represents $25$\% of the entire background, while the latter reaches $\approx40$\% of the total. 

We suggest, therefore, that assuming a non-unified AGN model, AGN are likely to account for at least 25\% of the total sub-mm background. But since this value is based on a model normalised by the ECDFS observations, where sub-mm sources counts are much lower than the average, the total AGN contribution may in fact be as high as 40\%.

However, assuming instead a unified AGN model the contribution is much smaller. In this case, the total predicted contribution is 5.9 Jy deg$^{-2}$, or $\approx13$\% of the total background. This provides an estimate of the lower limit on the AGN contribution to the EBL.

We note in passing that X-ray-undetected AGN in the ECDFS are predicted to contribute 9.7 Jy deg$^{-2}$ to the sub-mm background. This is to be compared to the measured $4.8 \pm 1.2$ Jy deg$^{-2}$ from the high-$z$ red galaxies. Again there is no inconsistency here, particularly given that the optical survey is flux-limited.

\section{Discussion} \label{s-disc}

We have found a strong correlation between X-ray sources and sub-mm sources. Our results indicate a significant AGN contribution to the sub-mm population, although it is not possible to claim immediately that the AGN are the main source of the bolometric output.

Sub-mm galaxies typically show strong PAH features (Lutz et al., 2005; Men\'endez-Delmestre et al., 2007; Pope et al., 2008; Coppin et al., 2010) and very high FIR luminosities (e.g.\ Alexander et al., 2005; Coppin et al., 2008b), thought to be fuelled by starbursts. These observations are generally taken to indicate that sub-mm galaxies are mostly dominated by star-fomation rather than AGN activity. However, strong PAH emission is observed even in sub-mm galaxies which are confirmed as AGN-dominated (Coppin et al., 2010), and Alexander et al. (2005) note that the FIR luminosities seen in sub-mm galaxies could arise from AGN if the dust covering factor is assumed to be large enough. 

Therefore, the question of whether AGN activity may be significant to the bolometric output of sub-mm sources is not entirely clear-cut. The strong correlation we have detected between 870\micron\ sources and X-ray sources continues to suggest that at least some of these sources host obscured AGN and the observations agree well with the predictions of the obscured AGN model. 

\emph{Furthermore, the bolometric luminosities of sub-mm galaxies} (see e.g.\ Chapman et al., 2005) \emph{are known to be extremely high} ($\ga10^{12}L_{\odot}$)\emph{, consistent with those of local ULIRGs and therefore also with quasars} (e.g.\ Sanders \& Mirabel, 1996). On the other hand, the bolometric luminosities of $z\approx2$ star-forming LBGs (Reddy et al., 2006) are at least an order of magnitude lower. If the most luminous star-forming galaxies are always heavily obscured then this may provide an explanation as to why these objects are not seen in the LBG population. Otherwise, the simplest interpretation may be that the sub-mm galaxies may contain the long-sought QSO population needed to explain the X-ray background.

We have considered the separate sub-mm contribution of X-ray sources at bright and faint fluxes and low and high levels of absorption. We have found some indication of a difference between absorbed and unabsorbed AGN, with the former preferentially being more submillimetre-bright. This is in line with the findings of other studies and supports a non-unified AGN model.

In the typical non-unified AGN model (Fabian, 1999), absorbed QSOs represent an earlier evolutionary phase in which there is rapid growth in the host spheroid, accompanied by a major episode of star formation. Unabsorbed quasars are seen during the subsequent phase, in which the host galaxy is quiescent. This is brought about by the expulsion of cool gas and dust from the galaxy once the spheroid is fully formed, a process which simultaneously quenches the star-formation, cuts off the sub-mm emission and renders the QSO unobscured. 

While this model invokes a star-formation episode, it is interesting to consider whether the AGN might itself be responsible for the sub-mm emission. Sub-mm emission is associated with dust at temperatures of a few tens of kelvin, therefore if AGN are submillimetre-bright they must maintain a cold dust component. Various authors using dust tori models to fit AGN data (e.g. Granato \& Danese, 1994; Andreani et al., 1999; Kuraszkiewicz et al., 2003) have shown that the dust surrounding the central nucleus in high-$z$ AGN can reach outer radii of $\sim1$ kpc, where the dust can be cold enough to produce sub-mm emission, with no requirement to invoke star-formation.

Modern approaches to the dust torus model tend to favour the idea of a `clumpy' torus, consisting of many separate clouds of gas and dust. The clumpiness of the torus could vary among AGN, with the more absorbed AGN having denser, dustier tori with greater optical depth. This would result in a lower intensity of radiation in the outer reaches of the torus, and therefore the dust at large radii would be cooler in absorbed AGN than in unabsorbed AGN.

\section{Summary} \label{s-summ}

We have used ECDFS data to test the contribution of AGN to the sub-mm number counts and background. We have used mainly statistical cross-correlation techniques which do not depend on the direct identification of any sub-mm source which is an advantage given the large beam size of the LABOCA sources. Our conclusions are as follows:

\begin{itemize}

\item We find a strong spatial correlation between 852 faint X-ray sources
and the 126 sub-mm sources. There is some suggestion that this correlation may be stronger for sources with 
faint X-ray fluxes and/or hard spectra, which tend to be associated with absorbed AGN.

\item We confirm that high luminosity X-ray sources show a stronger
correlation with the sub-mm than low-luminosity X-ray sources (Lutz et al., 2010). 
Since these high-$L_X$ sources also typically have faint fluxes, 
we conclude that they are at high redshifts, as expected, but also suggest they 
may largely be absorbed AGN, since we find evidence that 
$N_H>10^{22}$ cm$^{-2}$ sources dominate the correlation.

\item We find that the sub-mm sources also cross-correlate
with $z>0.5$ red galaxies, but not with bluer galaxy classes. These 
high-$z$ red galaxies have optical/NIR/FIR colours consistent with
those observed for X-ray sources.

\item The ECDFS X-ray sources to their flux limit make a contribution of
$1.5 \pm 0.1$ Jy deg$^{-2}$ to the sub-mm extragalactic background. 

\item We present a non-unified obscured AGN model which simultaneously
matches this observed sub-mm background, the bright ECDFS sub-mm
source counts and, significantly, the hard X-ray background.

\item In this model, the fractional AGN
contribution to the sub-mm background is at least 25\%, and
perhaps as high as $\approx 40$\%. The
remaining component is suggested to come from faint star-forming galaxies.

\item We find that in a unified AGN model the AGN contribution 
would be only $\approx13$\%. In this case the
unidentified bright sub-mm sources may be identified with obscured
starbursts rather than obscured AGN.

\item We measure the sub-mm background arising
from X-ray sources in different flux and luminosity ranges and find good agreement 
with the AGN model predictions in either the unified or non-unified case.

\item The sub-mm background contributions from absorbed ($N_H>10^{22}$ cm$^{-2}$)
and unabsorbed X-ray sources are also shown to agree well with the predictions of our non-unified AGN
model. In this case, however, we find that the predictions of the unified AGN model are poorly matched to the data.

\item The attraction of the non-unified model, therefore, is that it may simultaneously explain both 
the hard XRB and, at least, the bright sub-mm sources via a single obscured AGN
population. Other models not only need to invoke a new population of obscured,
ultraluminous star-forming galaxies to explain the bright sub-mm sources, but also
a fainter population of obscured AGN to explain the hard X-ray background.

\end{itemize}

\section*{Acknowledgements}

MDH acknowledges the receipt of an STFC studentship. We thank Ian Smail for providing access to the ECDFS LABOCA map and for useful discussions. Kristen Coppin and an anonymous referee are thanked for useful comments and suggestions. This work made use of data obtained with the APEX telescope, with programme IDs 078.F-9028(A), 079.F-9500(A), 080.A-3023(A) and 081.F-9500(A).

\label{lastpage}

\end{document}